\documentclass[
showpacs,  % Show PACS
showkeys,  % Show keywords
aps,       % APS style
amsthm,    % for math theorems
amsmath,   % for math formulas
amsfonts,  % for math fonts
amssymb    % for math symbols
]{revtex4-1}          % This is the latex document format corresponding to APS

\usepackage{amsthm}   % for math theorems
\usepackage{amsmath}  % for math formulas
\usepackage{amsfonts} % for math fonts
\usepackage{amssymb}  % for math symbols
\usepackage{graphicx} % Include figure files
\usepackage{dcolumn}  % Align table columns on decimal point
\usepackage{bm}       % Use boldsymbol
\usepackage{extarrows}% For example, extend the equal sign to write text
\usepackage[]{hyperref} % Add hypertext capabilities
\usepackage{float}    % Place figures
\usepackage{subfigure}% Many subfigures
\usepackage{epsfig}   % Use LaTex
\usepackage{color}    % Produce colored symbols
\usepackage[roman]{complexity}
\begin{document}
%\preprint{APS/123-QED}

\title{Quantum Speed Limit for Time-Fractional Open Systems}

\author{Dongmei Wei $^{1}$}

\author{Hailing Liu $^{1}$}

\author{Yongmei Li $^{1}$}

\author{Fei Gao $^{1}$}
\email{gaof@bupt.edu.cn}

\author{Sujuan Qin $^{1}$}

\author{Qiaoyan Wen $^{1}$}

\affiliation{$^{1}$ State Key Laboratory of Net\emph{}working and Switching Technology, Beijing University of Posts and Telecommunications, Beijing 100876, China}

\begin{abstract}
The Time-Fractional Schr\"{o}dinger Equation (TFSE) is well-adjusted to study a quantum system interacting with its dissipative environment. The Quantum Speed Limit (QSL) time captures the shortest time required for a quantum system to evolve between two states, which is significant for evaluating the maximum speed in quantum processes. In this work, we solve exactly for a generic time-fractional single qubit open system by applying the TFSE to a basic open quantum system model, namely the resonant dissipative Jaynes-Cummings (JC) model, and investigate the QSL time for the system. It is shown that the non-Markovian memory effects of the environment can accelerate the time-fractional quantum evolution, thus resulting in a smaller QSL time. Additionally, the condition for the acceleration evolution of the time-fractional open quantum system at a given driving time, i.e., a tradeoff among the fractional order, coupling strength and photon number, is brought to light. In particular, a method to manipulate the non-Markovian dissipative dynamics of a time-fractional open quantum system by adjusting the fractional order for a long driving time is presented.
\end{abstract}


\maketitle

\section{Introduction}
\label{intro}
The applications of fractional calculus to quantum processes have recently drawn extensive interest and played significant roles in describing nonlocal quantum phenomena \cite{Naber2004,Iomin2009,Ertik2010,Sirin2011,Achar2013,Iomin2019}. This quantum dynamics can be observed in the framework of the Time-Fractional Schr\"{o}dinger Equation (TFSE) \cite{Naber2004}, which is distinct from the Traditional Schr\"{o}dinger Equation (TSE) by the fractional-order derivative in time coordinate: $\frac{\partial }{{\partial t}} \rightarrow \frac{{{\partial ^\beta }}}{{\partial {t^\beta }}}$. Inspired by Laskin's work \cite{1Laskin2000,2Laskin2000}, Naber mapped the fractional Fokker-Planck equation into the TFSE and found the exact solutions of the TFSE for a free particle in a potential well. In particular, the time-fractional derivative in the TFSE, known as the Caputo fractional derivative in the form of a convolution integral with a power law memory kernel, is a flexible and appropriate mathematical tool to describe a quantum system coupled to its dissipative environment \cite{Tarasov2008,Iomin2009,Laskin2017}. Its extensions into the space-time-fractional quantum dynamics have been carried out, where the exact solutions for a free particle in an infinite square potential well and in a $\delta$-potential well were derived \cite{Wang2007,Dong2008}. Its physical relevance in a quantum comb model has been discussed and interpreted as a time evolution operator depicting the non-Markovian process of the quantum comb \cite{Iomin2009,Iomin2011}.

All quantum systems are open due to their inevitable interactions with an environment \cite{Breuer2002}, which leads to dissipation and/or decoherence since energy and/or information stream from the open system to the environment \cite{Breuer2002,Weiss2008}. The non-Markovian memory effects are induced by the backaction mechanisms resulting from the interaction \cite{Caruso2014}. To be specific, when energy and/or information stream back into the open system from the environment, the future states of the open system might be dependent on its past states, which suggests that there is nontrivial time correlativity among the states of the open system at different times in the whole dynamics \cite{Breuer2009,Fanchini2014,Addis2016}. Qubits, as we all know, serve as a fundamental element in new application fields such as quantum computation and communication, for which the interest in the theoretical analyses and actual implementations of two-level systems has been further stimulated. Several methods are available to create qubits with current quantum technologies, including quantum optics, microscopic quantum objects (electrons, ions, atoms) in traps, quantum dots and quantum circuits \cite{Majer2005,Berkley2003,Pashkin2003,Bellomo2007}. However, the different implementations of qubits are unavoidably subject to environmental noise. Therefore, the dissipative dynamics of two-level open systems have been a hot topic of research in open quantum system theory for recent years \cite{Maniscalco2006,Breuer2009,Deffner2013,Teittinen2019,Naikoo2020,Fanchini2022}.

With the booming field of quantum information science, it is a current need to achieve the rapid and controllable evolutions of quantum systems. Quantum Speed Limit (QSL) time, a fundamental bound on the evolution speed of a system imposed by the principles of quantum mechanics, is defined as the minimum time interval for the evolution of a quantum system between two separable states. For one side, it can be exploited to estimate the maximal speed of quantum simulations in almost all areas of quantum physics, which includes quantum information processing \cite{Lloyd2002}, quantum communication \cite{Lloyd2000}, quantum computing \cite{Bekenstein1981}, quantum metrology \cite{Giovanetti2011} and quantum entropy producing \cite{Deffner2010}. For another, it can help to evaluate the shortest time scales for convergence of quantum optimal control algorithms \cite{Hegerfeldt2013}, to optimize parameter measurement in metrology \cite{Braun2018}, to analyze the scrambling of spectral form factor \cite{Campo2017}, to manipulate decoherence \cite{Chenu2017}, to apply the reinforcement learning to seek the faster transfer schemes of quantum states \cite{Zhang2018}, etc. \cite{Fogarty2020,Deffner2020}. Initially, for closed quantum systems obeying unitary dynamics, the unified result of the QSL time can be acquired by combination of the Mandelstam-Tamm (MT) \cite{Mandelstam1991} and Margolus-Levitin (ML) bounds \cite{Margolus1998}. Since any quantum system inescapably couples to its environment, the lower bounds for the evolution time involving both the MT and ML types have been raised for open quantum systems with nonunitary evolutions \cite{ZhangYJ2014,Marvian2015,Campaioli2019}. Indeed, such speed limits can be deduced from classical systems as well \cite{Shanahan2018,Okuyama2018}.

Some impressive advances have been obtained in the analyses of the environmental influences on the QSL time of the traditional quantum dynamics over the years. It was discovered that the non-Markovian memory effects can raise the evolution speed of a single qubit open system by applying the TSE to the Jaynes-Cummings (JC) model for a two-level system coupled resonantly to a dissipative environment \cite{Deffner2013}, which had been validated in several circumstances \cite{XuZY2014,CaiXJ2017,XuK2018}. This non-Markovian environment-assisted acceleration was first experimentally confirmed in a cavity QED system \cite{Cimmarusti2015}. To our knowledge, no investigations on the QSL time for the time-fractional open quantum systems have emerged to date. Several questions then naturally occur: (i) is the acceleration phenomena, a rather common feature in the traditional quantum dynamics, also exist in the time-fractional quantum dynamics? (ii) can the non-Markovian memory effects speed up the time-fractional quantum evolution? (iii) what is the condition for accelerating the time-fractional quantum evolution in a memory environment?

For tackling these questions, in this peper, we work on the QSL time of the exact time-fractional quantum dynamics based on the resonant dissipative JC model which is to widely used to study the QSL time for open systems. It is demonstrated that the acceleration phenomena exist commonly not only in the traditional quantum dynamics ($\beta=1$), but also in the time-fractional one (including $\beta=1$). It is further shown that the non-Markovian memory effects of the environment can result in the faster time-fractional quantum evolution, and thus to the smaller QSL time. We discover that for a given driving time, the QSL time is related to three parameters: the fractional order of the Caputo fractional derivative, the coupling strength of the open system with its environment, and the photon number in a Fock state. In particular, the condition for speeding up the time-fractional quantum evolution in a memory environment is spelled out, that is, when both the coupling strength and the photon number are larger but the fractional order is smaller, the time-fractional quantum evolution will possess a greater capacity for potential speedup. In other words, it is the tradeoff among the three parameters that governs the further speedup evolution of the time-fractional open quantum system. It is remarkable that the crossover from the Markovian to non-Markovian processes can be observed in a long driving time when small fractional order is larger than a certain critical value. Therefore, by adjusting the fractional order, a method to manipulate the non-Markovian dissipative dynamics of the time-fractional open quantum system is presented.

The rest of this paper is structured as follows. In Sec. \ref{Sec:2}, we review the TFSE in Sec. \ref{Subsec:21}, the dissipative JC model in Sec. \ref{Subsec:22}, and derive the exact solutions for a generic time-fractional single qubit open system in Sec. \ref{Subsec:23}. Sec. \ref{Sec:3} presents the ML-type and ML-MT-type bounds of the QSL time applicable to open quantum systems, respectively. In Sec. \ref{Sec:4}, we analyze the identification and characterization of the non-Markovian quantum dynamics based on the resonant dissipative JC model in Sec. \ref{Subsec:41}, and investigate the condition for speeding up the time-fractional quantum evolution in a memory environment in Sec. \ref{Subsec:42}. The conclusion is reached in Sec. \ref{Sec:5}.

\section{Time-fractional open quantum systems}
\label{Sec:2}
In this section, we review the TFSE in Sec. \ref{Subsec:21} and the dissipative JC model in Sec. \ref{Subsec:22}. The exact solutions of a generic time-fractional single qubit open system will be derived in Sec. \ref{Subsec:23}.

\subsection{Time-Fractional Schr\"{o}dinger Equation (TFSE)}
\label{Subsec:21}
The time-fractional quantum dynamics can be described by the TFSE
\begin{equation}
\label{e1}
{(i\hbar )^\beta }\frac{{{\partial ^\beta }\left| {\psi (\boldsymbol{x},t)} \right\rangle }}{{\partial {t^\beta }}} = {H_\beta }\left| {\psi (\boldsymbol{x},t)} \right\rangle,
\end{equation}
where $0<\beta\leq1$ is the fractional order of the time-fractional derivative, ${H_\beta}$ is the pseudo-Hamiltonian of the time-fractional quantum system \cite{Dong2008}. To ensure the dimensional consistency in the time-fractional quantum dynamics, each variable and parameter is treated as dimensionless in Eq. (\ref{e1}), and thus $\hbar$ is also taken to be a dimensionless Planck constant \cite{Naber2004,Dong2008}. For $\beta=1$, Eq. (\ref{e1}) is the Traditional Schr\"{o}dinger Equation (TSE) with a first-order time derivative $\frac{\partial }{{\partial t}}$ to describe the traditional quantum dynamics. For $0<\beta<1$, the fractional-order time derivative $\frac{{{\partial ^\beta }}}{{\partial {t^\beta }}}$ is formed by a convoluted integration with a power law memory kernel \cite{Iomin2009}
\begin{equation}
\label{e2}
\frac{{{\partial ^\beta }\left| {\psi (t)} \right\rangle }}{{\partial {t^\beta }}} \equiv D_t^\beta \left| {\psi (t)} \right\rangle  = \int_0^t {\frac{{{{(t - \tau )}^{ - \beta }}}}{{\Gamma (1 - \beta )}}} \frac{{\partial \left| {\psi (t)} \right\rangle }}{{\partial t}}d\tau  = k(t) * \left| {\dot \psi (t)} \right\rangle,
\end{equation}
which is defined as the Caputo fractional derivative \cite{Iomin2009}, $\Gamma (z) = \int_0^{ + \infty } {{t^{z - 1}}} {e^{ - t}}dt\,(z \in C)$ is the gamma function, and $k(t) = \frac{{{{(t - \tau )}^{ - \beta }}}}{{\Gamma (1 - \beta )}}$ is the memory kernel function. The definition is of great interest for solving the TFSE, both to make it feasible to apply the Laplace transform to the time-fractional derivative and to provide a time-fractional evolution supported by the Mittag-Leffler function. If ${H_\beta}$ is time-independent, the wave function with a time-fractional evolution operator can be given by \cite{Iomin2009}
\begin{equation}
\label{e3}
\left| {\psi (t)} \right\rangle  = {E_\beta }\left[ {{{( - it)}^\beta }{H_\beta }} \right]\left| {\psi (0)} \right\rangle,
\end{equation}
where ${E_{\beta ,1}}(z) \equiv {E_\beta }(z) = \int_{j = 0}^{ + \infty } {\frac{{{z^j}}}{{\Gamma (\beta j + 1)}}}$ is the Mittag-Leffler function. Interestingly, the time evolution operator ${E_\beta}$ can be split into the oscillation and decay part in time due to the path integration combined with the residue contributions \cite{Naber2004,Iomin2009}, represented as follows
\begin{equation}
\label{e4}
{E_\beta }\left[ {{{(\frac{{ - it}}{\hbar })}^\beta }{H_\beta }} \right] = \frac{{{e^{\frac{{ - i{H_\beta }^{1/\beta }t}}{\hbar }}}}}{\beta } - \frac{{{H_\beta }{i^\beta }{\hbar ^\beta }\sin (\beta \pi )}}{\pi }\int_0^{ + \infty } {\frac{{{e^{ - rt}}{r^{\beta  - 1}}dr}}{{{r^{2\beta }} - 2{H_\beta }{i^\beta }{\hbar ^\beta }\cos (\beta \pi ){r^\beta } + {{({H_\beta }{i^\beta }{\hbar ^\beta })}^2}}}}.
\end{equation}
Clearly, the time-fractional approach can be used for the description of the time evolution of an action on the basis of the theories of the evolution operator in mathematics and the time evolution in physics. It aims to depict the dynamics of a quantum system through the history information depending on the changes of time intervals, rather than depicting that according to the changes of point times.

\subsection{Dissipative Jaynes-Cummings (JC) model}
\label{Subsec:22}
The dissipative JC model, a interesting microscopic open quantum system model, is capable of characterizing many different physical systems to provide a solid understanding of the physical phenomena associated with the non-Markovian memory effects of the environment \cite{Garraway1997}.

In the framework of the rotating wave approximation, the total Hamiltonian ($\hbar=1$) that describes a two-level atom (qubit system) interacting with a one-mode leaky cavity (dissipative environment) at zero temperature can be formulated in the following form
\begin{equation}
\label{e5}
{H^{_{total}}} = {H^S} + {H^E} + {H^I}.
\end{equation}
Here
\begin{equation}
\label{e6}
{H^S} = {\omega _0}{\sigma _ + }{\sigma _ - }
\end{equation}
and
\begin{equation}
\label{e7}
{H^E} = \sum\limits_j {{\omega _j}} b_j^\dag {b_j}
\end{equation}
are the free Hamiltonian of the qubit as well as that of the environment, respectively. ${\omega _0}$ refers to the transition frequency of the two-level atom over the excited and ground states. ${\sigma _ \pm }$ stand for the atomic inversion operators. ${\omega _j}$ and $b_j^\dag$ (${b_j}$) are correspondingly the angular frequency and the creation (annihilation) operators of the $j$th mode of the cavity.

The interaction Hamiltonian is
\begin{equation}
\label{e8}
{H^I} = \sum\limits_j {{\lambda _j}(} b_j^\dag {\sigma _ - } + {b_j}{\sigma _ + })
\end{equation}
between the two-level atom system and its cavity field environment with the coupling strength ${\lambda _j}\in\left[{0,1}\right]$. Since now, we will concentrate on the interaction picture defined by ${H^S} + {H^E}$, where the interaction Hamiltonian is rewritten to
\begin{equation}
\label{e9}
{H^I}(t) = \sum\limits_j {{\lambda _j}(} b_j^\dag {\sigma _ - }{e^{ - i({\omega _0} - {\omega _j})t}} + {b_j}{\sigma _ + }{e^{i({\omega _0} - {\omega _j})t}}).
\end{equation}

\subsection{Exact solutions for a generic time-fractional single qubit open system}
\label{Subsec:23}
We will work out the exact solutions for a generic time-fractional single qubit open system by applying the TFSE to the resonant dissipative JC model. Here, one excitation is assumed to exist in the composite atom-cavity system, in which the excited and ground states of the atom are respectively $\left| e \right\rangle$ and $\left| g \right\rangle$, and the cavity in the form of single mode (i.e., $j=1$) is initially at the Fock state denoted by $\left| n \right\rangle$ ($n\in{N}$). Based on the above results, ${H^I}(t)$ can only lead to the transition dominating the atom-cavity dynamics to be $\left| {e,n} \right\rangle  \leftrightarrow \left| {g,n + 1} \right\rangle$. Therefore, the density matrix of ${H^I}(t)$ in the composite computational basis $\left\{ {\left| {g,n + 1} \right\rangle, \left| {e,n} \right\rangle } \right\}$ can be expressed as
\begin{equation}
\label{e10}
{H^I}(t) = \left( {\begin{array}{*{20}{c}}
0&{\lambda \sqrt {n + 1} {e^{ - i\Delta t}}}\\
{\lambda \sqrt {n + 1} {e^{i\Delta t}}}&0
\end{array}} \right),
\end{equation}
where $\Delta  = {\omega _0} - {\omega _j}$ means the detuning. In the case of $\Delta>0$, the Hamiltonian is non-Hermitian with respect to time. When the atomic transition resonates with the cavity mode, that is, $\Delta=0$, the density matrix of ${H^I}(t)$ can further be shown in terms of the time-independent Hermitian operator
\begin{equation}
\label{e11}
{H^I} = \left( {\begin{array}{*{20}{c}}
0&{\lambda \sqrt {n + 1} }\\
{\lambda \sqrt {n + 1} }&0
\end{array}} \right).
\end{equation}

Since ${H^I}$ is a Hermitian operator, it has a spectral decomposition which can be expressed by ${H^I} = \sum\limits_i {{\alpha _i}} \left| {{u_i}} \right\rangle \left\langle {{u_i}} \right|$ with the eigenvalues ${{\alpha _i}}$ and the corresponding normalized eigenvectors $\left| {{u_i}} \right\rangle$. By substituting the spectral decomposition of ${H^I}$ into Eq. (\ref{e3}), we can obtain the exact solutions to the TFSE for the combined atom-cavity system as displayed below
\begin{eqnarray}
\label{e12}
\begin{array}{l}
{\left| {\psi (t)} \right\rangle _{SE}} = {E_\beta }\left[ {{{( - it)}^\beta }H_\beta ^I} \right]{\left| {\psi (0)} \right\rangle _{SE}}\\
{\kern 1pt} {\kern 1pt} {\kern 1pt} {\kern 1pt} {\kern 1pt} {\kern 1pt} {\kern 1pt} {\kern 1pt} {\kern 1pt} {\kern 1pt} {\kern 1pt} {\kern 1pt} {\kern 1pt} {\kern 1pt} {\kern 1pt} {\kern 1pt} {\kern 1pt} {\kern 1pt} {\kern 1pt} {\kern 1pt} {\kern 1pt} {\kern 1pt} {\kern 1pt} {\kern 1pt} {\kern 1pt} {\kern 1pt} {\kern 1pt} {\kern 1pt} {\kern 1pt} {\kern 1pt} {\kern 1pt} {\kern 1pt} {\kern 1pt} {\kern 1pt} {\kern 1pt} {\kern 1pt} {\kern 1pt} {\kern 1pt} {\kern 1pt} {\kern 1pt} {\kern 1pt}  = \left[ {\sum\limits_i {{E_\beta }\left[ {{{( - it)}^\beta }{\alpha _i}} \right]{{\left| {{u_i}} \right\rangle }_{SE}}\left\langle {{u_i}} \right|} } \right]{\left| {\psi (0)} \right\rangle _{SE}}\\
{\kern 1pt} {\kern 1pt} {\kern 1pt} {\kern 1pt} {\kern 1pt} {\kern 1pt} {\kern 1pt} {\kern 1pt} {\kern 1pt} {\kern 1pt} {\kern 1pt} {\kern 1pt} {\kern 1pt} {\kern 1pt} {\kern 1pt} {\kern 1pt} {\kern 1pt} {\kern 1pt} {\kern 1pt} {\kern 1pt} {\kern 1pt} {\kern 1pt} {\kern 1pt} {\kern 1pt} {\kern 1pt} {\kern 1pt} {\kern 1pt} {\kern 1pt} {\kern 1pt} {\kern 1pt} {\kern 1pt} {\kern 1pt} {\kern 1pt} {\kern 1pt} {\kern 1pt} {\kern 1pt} {\kern 1pt} {\kern 1pt} {\kern 1pt} {\kern 1pt} {\kern 1pt}  = \left[ \begin{array}{l}
E[{( - it)^\beta }(\lambda \sqrt {n + 1} )]{\kern 1pt} \\
{\kern 1pt} (a{\left| {g,n + 1} \right\rangle _{SE}} + b{\left| {e,n} \right\rangle _{SE}}){\kern 1pt} (a{\left\langle {g,n + 1} \right|_{SE}} + b{\left\langle {e,n} \right|_{SE}}){\kern 1pt} \\
{\kern 1pt}  + E[{( - it)^\beta }( - \lambda \sqrt {n + 1} )]\\
{\kern 1pt} ( - a{\left| {g,n + 1} \right\rangle _{SE}} + b{\left| {e,n} \right\rangle _{SE}})(-a{\left\langle {g,n + 1} \right|_{SE}} + b{\left\langle {e,n} \right|_{SE}})
\end{array} \right]\,{\left| {\psi (0)} \right\rangle _{SE}}
\end{array}
\end{eqnarray}
with $a,b\in[0,1]$ and the normalized condition $a^2+b^2=1$ for the eigenvectors. The physical system of concern is the two-level atom system obeying time-fractional evolution. With the initial state of the composite system known, the reduced density matrix (i.e., the solutions of the time-fractional open quantum system) that plays a really effective role after tracing out the degrees of freedom in the cavity environment can be exactly calculated to ${\rho _S} = t{r_E}({\left| {\psi (t)} \right\rangle _{SE}}\left\langle {\psi (t)} \right|)$.

\section{Quantum speed limit time}
\label{Sec:3}
In order to explore the evolution speed of the time-fractional open quantum system, we should proceed from the definition of the QSL time applicable to open quantum systems. The QSL time provides an effective definition on a lower bound of the evolution time for any initial state, which contributes to find the fastest evolution speed of the open system. In this section, we introduce two different types of bounds for the QSL time serving an open system, including ML-type and ML-MT-type bounds.

\subsection{ML-type bound}
\label{Subsec:31}
A unified lower ML-type bound on the QSL time for any driven open system was introduced by Deffner and Lutz \cite{Deffner2013}. Taking advantage of von Neumann trace inequality and Cauchy-Schwarz inequality, the QSL time from an initial pure state ${\rho _S}(0) = {\left| {\psi (0)} \right\rangle _S}\left\langle {\psi (0)} \right|$ of the open system to its target state ${\rho _S}(\tau)$ can be obtained by
\begin{equation}
\label{e13}
{\tau _{QSL}} = \max \left\{ {\frac{1}{{\Lambda _\tau ^{tr}}},\frac{1}{{\Lambda _\tau ^{hs}}},\frac{1}{{\Lambda _\tau ^{op}}}} \right\}{\sin ^2}\left[ {B({\rho _S}(0),{\rho _S}(\tau ))} \right]
\end{equation}
with a Bures angle $B({\rho _S}(0),{\rho _S}(\tau )) = \arccos \sqrt {\left\langle {\psi (0)} \right|{\rho _S}(\tau )\left| {\psi (0)} \right\rangle }$ between the initial and target states of the open system. $\Lambda _\tau ^p = \frac{1}{\tau }{\int_0^\tau  {\left\| {{{\dot \rho }_S}(t )} \right\|} _p}dt$ means the average of ${\left\| {{{\dot \rho }_S}(t )} \right\|_p}$ on the driving time $\tau  \in \left[ {0,1} \right]$, and $\,{\left\| A \right\|_p} = {(s _1^p + s _2^p +  \cdots  + s _m^p)^{{1 \mathord{\left/{\vphantom {1 p}} \right.\kern-\nulldelimiterspace} p}}}$ is the Schatten $p$-norm, where $p$ is either $p=tr$, $p=hs$ or $p=op$ for trace, Hilbert-Schmidt and operator norm, respectively. With the above definitions, $\,{\left\| A \right\|_p}$ can clearly be displayed as follows
\begin{equation}
\label{e14}
{\left\| A \right\|_{tr}} = \sum\limits_m {{s_m}} ,\,\,\,\,\,\,\,\,\,{\left\| A \right\|_{hs}} = \sqrt {\sum\limits_m {s_m^2} } ,\,\,\,\,\,\,\,\,\,{\left\| A \right\|_{op}} = \mathop {\max }\limits_m \left\{ {{s_m}} \right\},
\end{equation}
where ${s _1},{s _2},\cdots,{s _m}$ are the singular values of $A$. It was finally shown that the ML-type bound dependent on the operator norm of the nonunitary generator provided a sharper and tighter bound on the QSL time than any bound \cite{Deffner2013}, and the QSL time could accordingly be simplified to be
\begin{equation}
\label{e15}
{\tau _{QSL}} = \frac{{{{\sin }^2}\left[ {B({\rho _S}(0),{\rho _S}(\tau ))} \right]\,}}{{\Lambda _\tau ^{op}}}\,.
\end{equation}

\subsection{ML-MT-type bound}
\label{Subsec:32}
To be available for both pure and mixed initial states, a general ML-MT-type (i.e., including both ML and MT types) bound on the QSL time for any driven open system has been investigated by Zhang et al. \cite{ZhangYJ2014}. With the use of the relative purity as well as von Neumann trace inequality and Cauchy-Schwarz inequality, the QSL time for any initial state of the open system is read with
\begin{equation}
\label{e16}
{\tau _{QSL}} = \,\max \left\{ {\frac{1}{{\overline {\sum\limits_m {{s_m}{v_m}} } }},\frac{1}{{\overline {\sqrt {\sum\limits_m {s_m^2} } } }}} \right\}\left| {f(\tau  + {\tau _D}) - 1} \right|tr({\chi _S}{(\tau )^2}),
\end{equation}
where $\overline Z  = \frac{1}{{{\tau _D}}}\int_\tau ^{\tau  + {\tau _D}} {Zdt}$ denotes the average of $Z$ at the driving time $\tau _D$, ${{s_m}}$ and ${{v_m}}$ are the singular values of ${{\dot \chi }_S}(t)$ and the initial mixed state ${\chi _S}(\tau )$, respectively. $f(\tau  + {\tau _D}) = {{tr\left[ {{\chi _{\tau  + {\tau _D}}}{\chi _\tau }} \right]} \mathord{\left/{\vphantom {{tr\left[ {{\chi _{\tau  + {\tau _D}}}{\chi _\tau }} \right]} {tr(\chi _\tau ^2)}}} \right.\kern-\nulldelimiterspace} {tr(\chi _\tau ^2)}}$ defines the relative purity between the initial state $\chi_\tau$ and its target state $\chi_{\tau+{\tau _D}}$.

One of the significant applications of the QSL time is to assess the evolution speed of a quantum system under two cases as follows \cite{Liuchen2015}: (i) ${\tau _{QSL}}/{\tau}=1$ shows that the quantum system has evolved at the fastest speed when the actual driving time $\tau$ reaches the QSL time $\tau_{QSL}$, that is, no acceleration phenomena in the quantum evolution will occur. Hence, no potential ability to accelerate the quantum evolution will exist. (ii) In the case of ${\tau _{QSL}}/{\tau}<1$, the quantum system undergoes a slower evolution and its acceleration evolution may take place. If the ratio ${\tau _{QSL}}/{\tau}$ is lower, the potential ability to accelerate the quantum evolution will be greater.

\section{Non-Markovian dissipative dynamics of time-fractional open quantum systems}
\label{Sec:4}
In this section, the identification and characterization of the non-Markovian quantum dynamics based on the resonant dissipative JC model are analyzed in Sec. \ref{Subsec:41}. The acceleration condition in the evolution of the time-fractional open quantum system is investigated in Sec. \ref{Subsec:42}.

\subsection{Non-Markovian quantum dissipative dynamics}
\label{Subsec:41}
\begin{figure}
\includegraphics[width=0.99\textwidth]{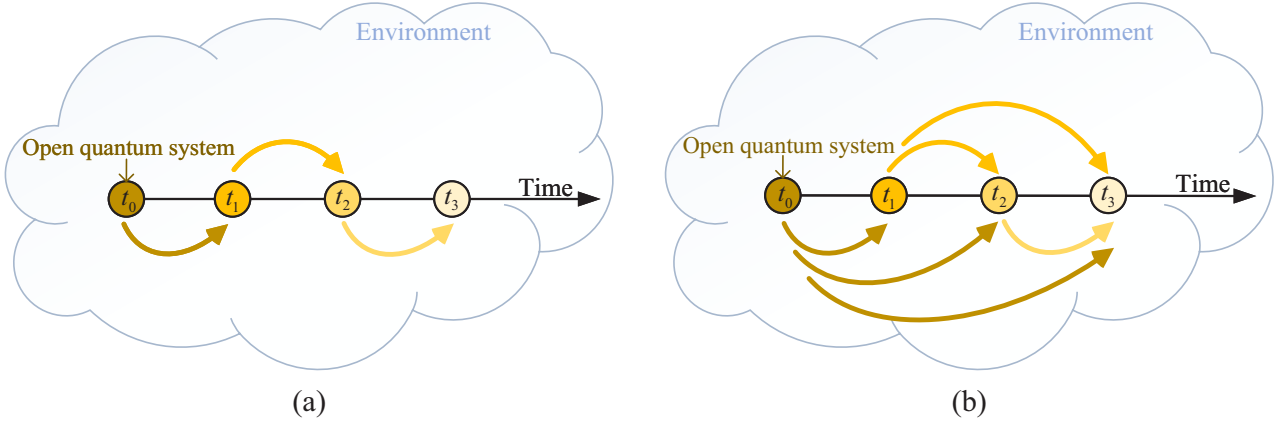}
\caption{Non-Markovian quantum dissipative dynamics. (a) Schematic illustration of the trajectory describing the Markovian dynamics of an open quantum system. For this case, the state of the open system (dots) at time ${t_{\eta  + 1}}$ is only dependent on that at time of its immediate preceding step, that is, at time ${t_\eta }$. (b) Schematic illustration of the time evolution of an open quantum system in a non-Markovian manner. Under the circumstance, the state at time ${t_{\eta  + 1}}$ is generally dependent on the states of the open system throughout the trajectory.}
\label{Fig1}
\end{figure}

In the resonant dissipative JC model under consideration, the interacting of the quantum system with its dissipative environment gives rise to dissipation and decoherence in the flows of energy and information from the open system to the environment. The emergence of the non-Markovian memory effects has a close relation with the dynamical exchanges of energy and information between the open system and the environment. When the backflow of energy and information from the environment to the open system takes place, the non-Markovian memory effects emerge throughout the open system dynamics. Based on the above, a schematic illustration of the trajectory identifying the non-Markovian quantum dissipative dynamics is presented in Fig. \ref{Fig1}.

The characterization of the non-Markovian quantum dissipative dynamics has been a part of great significance in the study of open quantum systems. Interestingly, the distinct non-Markovian features have been observed in the explorations of the traditional quantum dynamics based on the resonant dissipative JC model. For instance, the exact dynamic of the exited state population showed the damped oscillation behaviors \cite{Maniscalco2006,Ferraro2009}. The entanglement dynamics for any initial state vanished and revived periodically, with a damped amplitude for its revival \cite{Bellomo2007,Addis2016}. The pairwise oscillating translation and deformation of the QSL time for a general single qubit open dynamic were found in a non-Markovian region \cite{Deffner2013,Teittinen2019}, etc. \cite{XuZY2014,CaiXJ2017,Naikoo2020}. Such pairs of oscillations and decays induced by the non-Markovian memory effects of the dissipative environment are described as the non-Markovian oscillations, which are concretely manifested as collapses and partial revivals. In a physical sense, these non-Markovian oscillations correspond to the dynamical exchanges of energy and information between the open system and the environment by reason of the environmental memory.

Throughout the time-fractional quantum dynamics, we now know, the past history of a quantum system are explicitly considered by employing the the Caputo fractional derivative. Its physical relevance has been interpreted as a time evolution operator, which can depict the valid interaction between an open quantum system and its dissipative environment \cite{Iomin2009}. It is appropriate to stress that the time evolution operator is composed of the oscillatory and decay parts. Therefore, by applying the TFSE to the resonant dissipative JC model, the non-Markovian oscillations in the time evolution of the time-fractional open quantum system can be well displayed in this paper.

\subsection{Condition for the speedup evolution of the time-fractional open quantum system}
\label{Subsec:42}
We employ the ML-type bound to calculate the QSL time for a generic time-fractional single qubit open system from an initial pure state ${\rho _S}(0)$ to its target state ${\rho _S}(\tau )$ through a driving time $\tau$. For this purpose, it is assumed that a two-level atom (qubit system) in the excited state with one excitation and a one-mode leaky cavity (dissipative environment) in the Fock state are prepared at $\tau=0$,
\begin{equation}
\label{e17}
{\left| {\psi (0)} \right\rangle _{SE}} = {\left| e \right\rangle _S}{\left| n \right\rangle _E}.
\end{equation}
Substituting Eq. (\ref{e11}) and Eq. (\ref{e17}) into Eq. (\ref{e12}) then the combined target state at the driving time $\tau$ can be gained by
\begin{equation}
\label{e18}
\begin{aligned}
{\left| {\psi (\tau )} \right\rangle _{SE}} &= ab({E_{{\beta _2}}} - {E_{{\beta _1}}}){\left| {g,n + 1} \right\rangle _{SE}}\\
&+{b^2}({E_{{\beta _2}}} + {E_{{\beta _1}}}){\left| {e,n} \right\rangle _{SE}},
\end{aligned}
\end{equation}
where
\begin{equation}
\label{e19}
\begin{aligned}
{E_{{\beta _1}}} &= {E_\beta }\left[ {{{( - i\tau )}^\beta }( - \lambda )\sqrt {n + 1}} \right],\\
{E_{{\beta _2}}} &= {E_\beta }\left[ {{{( - i\tau )}^\beta }\lambda \sqrt {n + 1} } \right].
\end{aligned}
\end{equation}

After tracing out the degrees of freedom of the cavity environment, the reduced density matrix of the atom system obeying time-fractional evolution takes the exact form in the standard computational basis $\left\{ {\left| g \right\rangle,\left| e \right\rangle } \right\}$
\begin{eqnarray}
\label{e20}
\begin{aligned}
{\rho _S}(\tau ) &= t{r_E}({\left| {\psi (\tau )} \right\rangle _{SE}}\left\langle {\psi (\tau )} \right|)\\
\,\,\,\,\,\,\,\,\,\,\,\,\, &= \frac{1}{N}\left( {\begin{array}{*{20}{c}}
{{a^2}{b^2}{{\left| {\,{E_{{\beta _2}}} - {E_{{\beta _1}}}} \right|}^2}}&0\\
0&{{b^4}{{\left| {\,{E_{{\beta _2}}} + {E_{{\beta _1}}}} \right|}^2}}
\end{array}} \right),
\end{aligned}
\end{eqnarray}
where $N = {a^2}{b^2}{\left| {\,{E_{{\beta _2}}} - {E_{{\beta _1}}}} \right|^2} + {b^4}{\left| {\,{E_{{\beta _2}}} + {E_{{\beta _1}}}} \right|^2}$ is the normalization factor. According to Eq. (\ref{e20}), we can first have
\begin{eqnarray}
\label{e21}
\begin{aligned}
{\sin ^2}\left[ {B({\rho _S}(0),{\rho _S}(\tau ))} \right] &= \left| {tr({\rho _S}(0){\rho _S}(\tau )) - 1} \right|\\
&= \left| {\frac{{{a^2}({E_{{\beta _1}}}E_{{\beta _1}}^ *  - {E_{{\beta _2}}}E_{{\beta _1}}^ *  - {E_{{\beta _1}}}E_{{\beta _2}}^ *  + {E_{{\beta _2}}}E_{{\beta _2}}^ * )}}{{({b^2} - {a^2})({E_{{\beta _1}}}E_{{\beta _2}}^ *  + {E_{{\beta _2}}}E_{{\beta _1}}^ * ) - ({b^2}+{a^2})({E_{{\beta _1}}}E_{{\beta _1}}^ *  + {E_{{\beta _2}}}E_{{\beta _2}}^ * )}}}\right|.
\end{aligned}
\end{eqnarray}
Then our next tasks are to work out all singular values of ${\dot \rho _S}(t)$ and locate the maximum singular value ${s_{\max }} = \,{\left\| {{{\dot \rho }_S}(t)\,} \right\|_{op}}$. Thus, for a generic time-fractional single qubit open system, the QSL time expressed in Eq. (\ref{e13}) is ultimately written to be
\begin{eqnarray}
\label{e22}
\frac{{{\tau _{QSL}}}}{\tau } = \frac{{\left| {\frac{{a^2({E_{{\beta _1}}}E_{{\beta _1}}^ *  - {E_{{\beta _2}}}E_{{\beta _1}}^ *  - {E_{{\beta _1}}}E_{{\beta _2}}^ *  + {E_{{\beta _2}}}E_{{\beta _2}}^ * )}}{{({b^2} - {a^2})({E_{{\beta _1}}}E_{{\beta _2}}^ *  + {E_{{\beta _2}}}E_{{\beta _1}}^ * )-({b^2}+{a^2})({E_{{\beta _1}}}E_{{\beta _1}}^ *  + {E_{{\beta _2}}}E_{{\beta _2}}^ * )}}} \right|}}{{\int_0^\tau  {{s_{\max }}} dt}}.
\end{eqnarray}
Obviously, apart from the state parameters $a, b$ and the driving time $\tau$, the QSL time here is also related to three parameters: the fractional order $\beta$, the coupling strength $\lambda$ and the photon number $n$. For convenience, we choose $a=b=\sqrt{2}/2$. Now we numerically examine the influences of $\beta$, $\lambda$ and $n$ on the QSL time in the framework of the time-fractional open quantum system.

\begin{figure}
\includegraphics[width=0.8\textwidth]{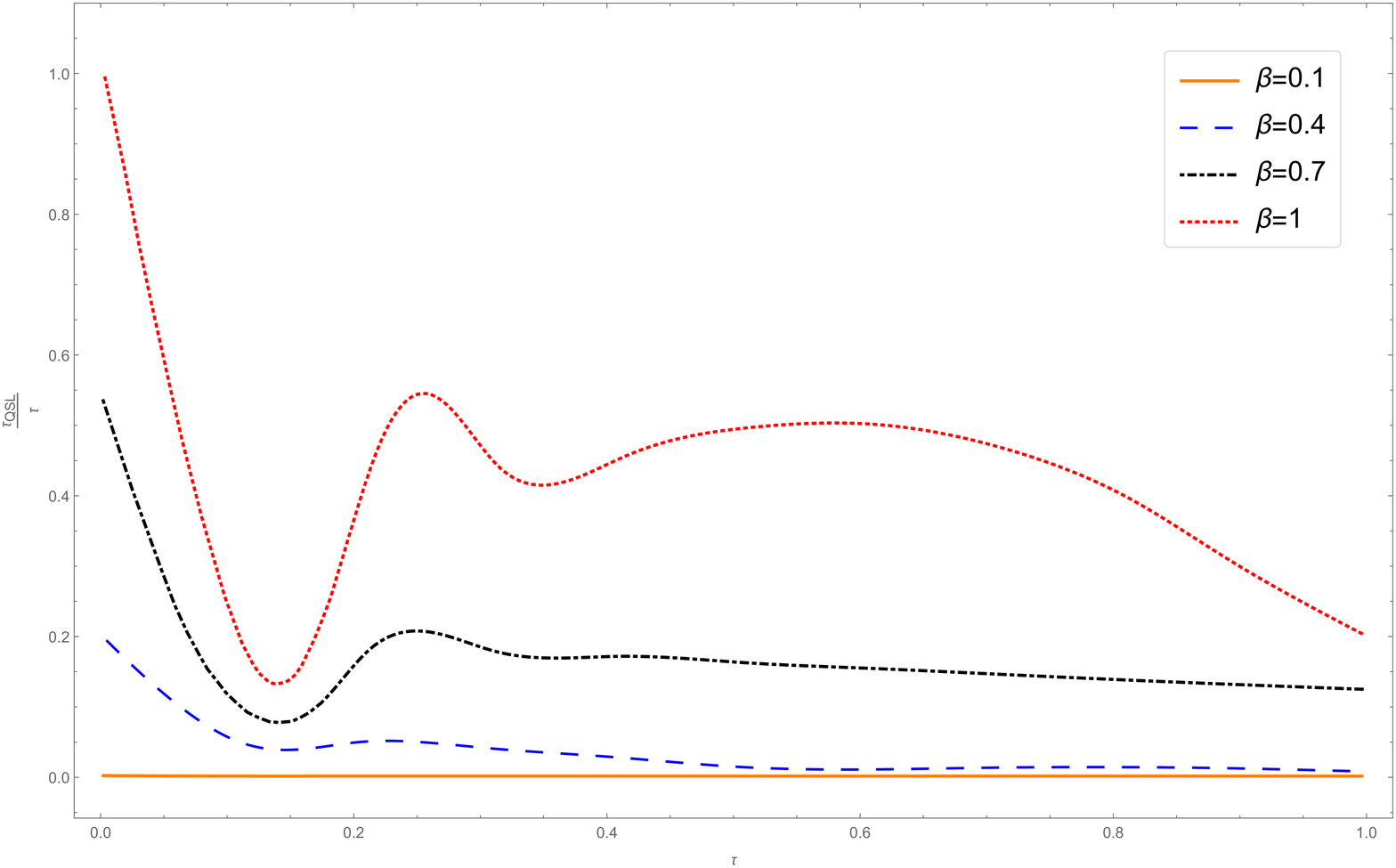}
\caption{QSL time for a generic time-fractional single qubit open system, quantized by the ratio ${\tau _{QSL}}/{\tau}$ as a function of the driving time $\tau$, for different fractional order $\beta=0.1, 0.4, 0.7, 1$. Parameters are chosen as $\lambda = 0.5$ and $n=20$.}
\label{Fig2}
\end{figure}

In Fig. \ref{Fig2}, we present the QSL time for a generic time-fractional single qubit open system. It is quantized by the ratio ${\tau _{QSL}}/{\tau}$ in Eq. (\ref{e22}) as a function of $\tau$ for different choices of $\beta=0.1, 0.4, 0.7, 1$ with $\lambda = 0.5$, $n=20$. We find that the acceleration phenomena are as common in the time-fractional quantum dynamics ($0<\beta\leq1$) as in the traditional one ($\beta=1$). Moreover, the QSL time is notably dependent on $\beta$, and behaves as the non-Markovian dissipative dynamics when small $\beta$ is larger than a certain critical value. Hence, the transition from the Markovian to non-Markovian processes can be manipulated by adjusting $\beta$. We also note that as $\beta$ gets larger, the frequencies and amplitudes of the non-Markovian oscillations are stronger. And when $\beta$ is large enough, these non-Markovian oscillations may still be remained as $\tau$ increases, which indicates that the non-Markovian dissipative dynamics of the time-fractional open quantum system may be validly described in long driving times. As a note, for the time-fractional open quantum system, the smaller $\beta$ has a beneficial effect on reducing the QSL time, that is, the smaller $\beta$ can yield the more capacity for potential acceleration evolution.

\begin{figure}[htbp]
\vspace{0.5cm}
\subfigtopskip=2pt
\subfigbottomskip=5pt
\subfigcapskip=-5pt
\centering
\subfigure[$\beta$=0.2]{
    \label{Fig3(a)}
    \includegraphics[width=0.46\linewidth]{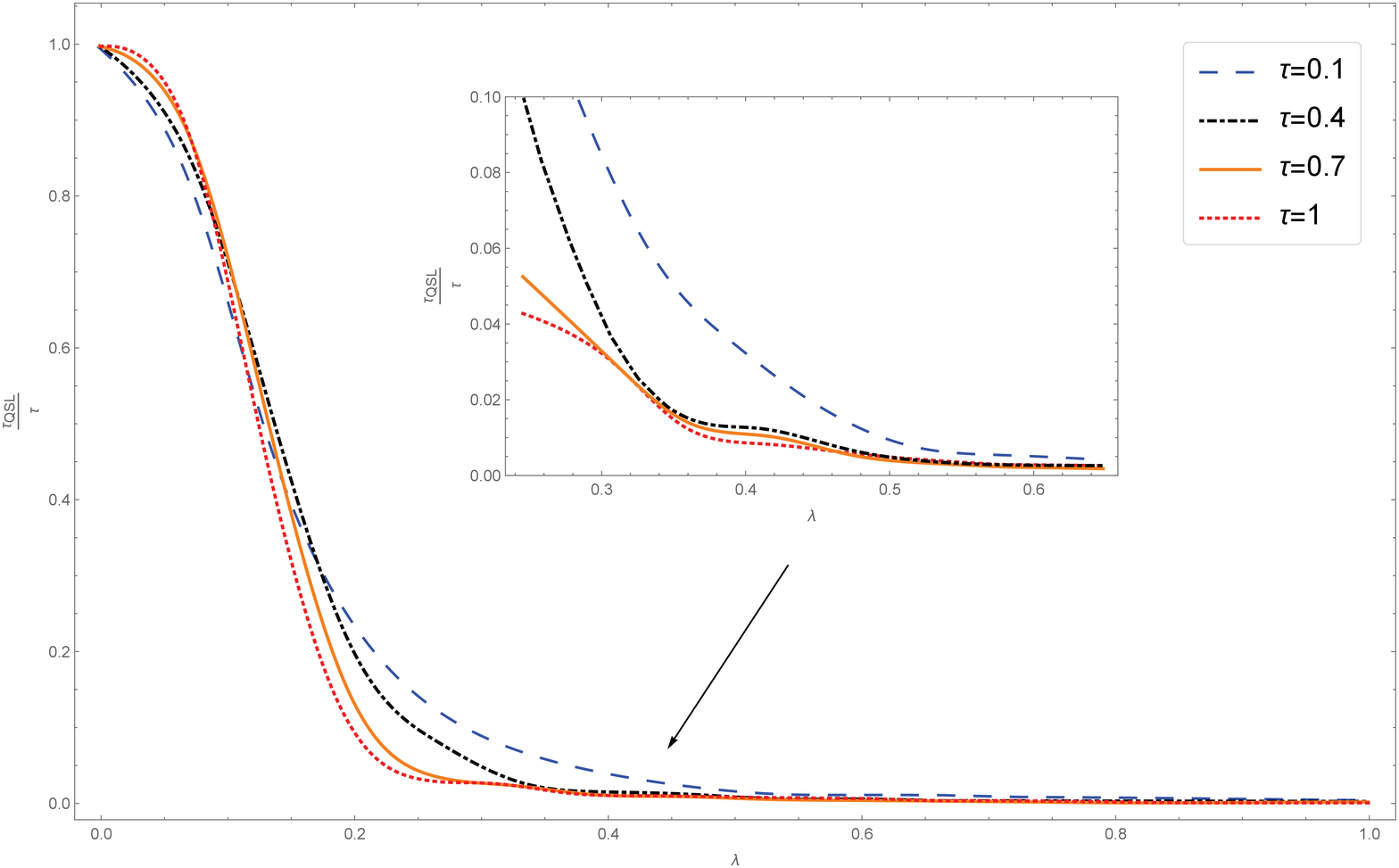}}
    \quad
\subfigure[$\beta$=0.5]{
    \label{Fig3(b)}
    \includegraphics[width=0.46\linewidth]{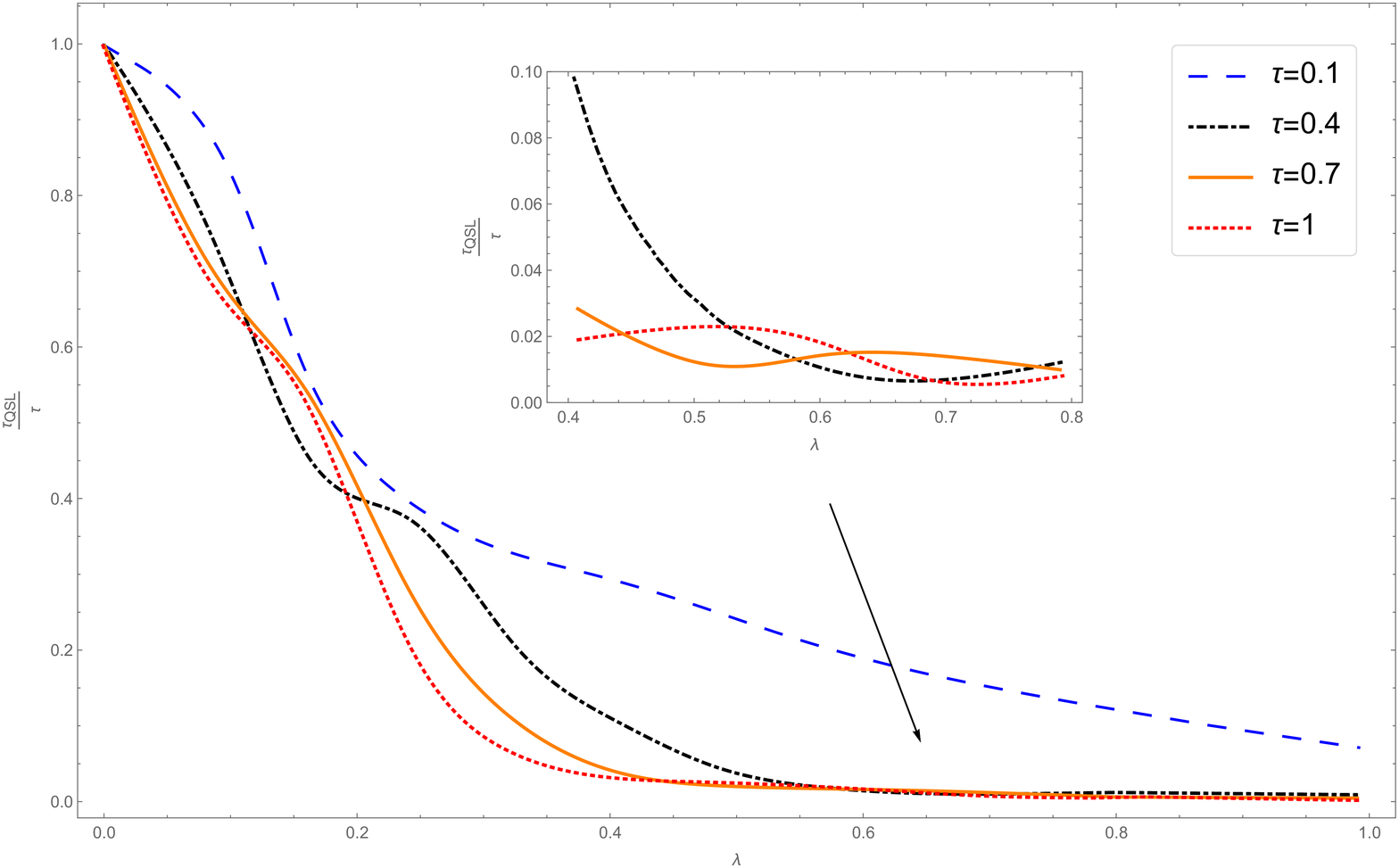}}\\
\subfigure[$\beta$=0.8]{
    \label{Fig3(c)}
    \includegraphics[width=0.46\linewidth]{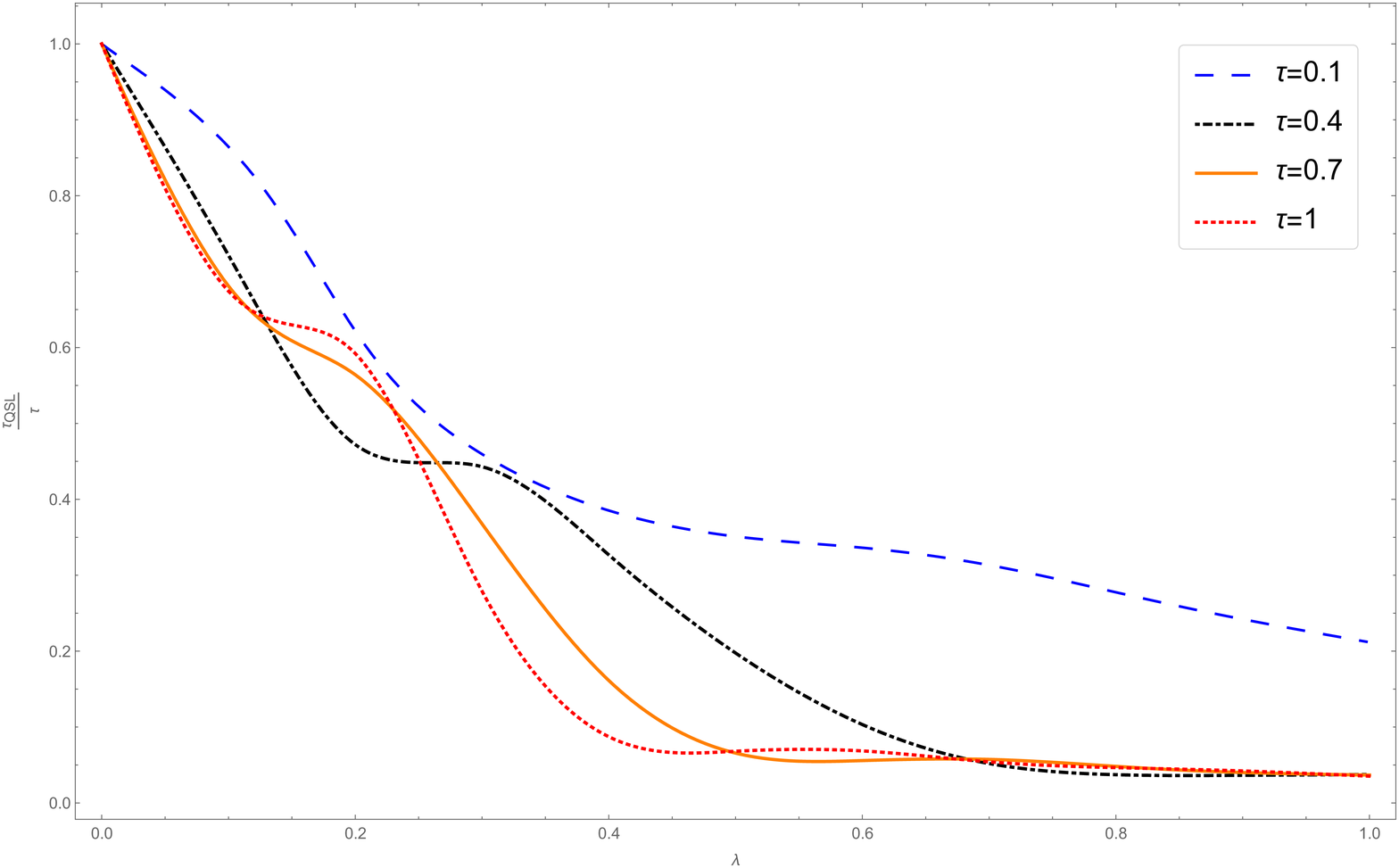}}
    \quad
\subfigure[$\beta$=1]{
    \label{Fig3(d)}
    \includegraphics[width=0.46\linewidth]{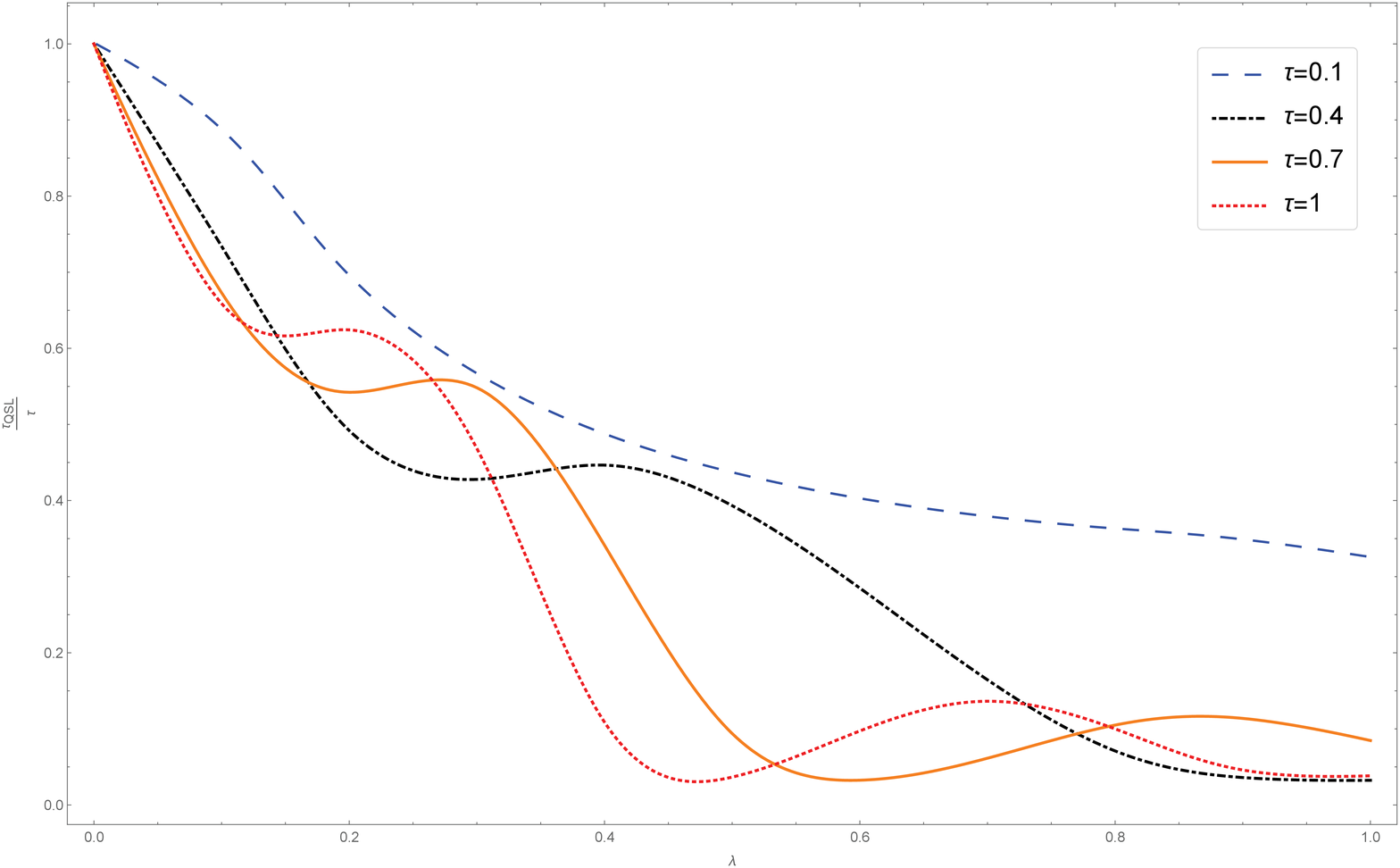}}\\
\caption{Comparison of ${\tau _{QSL}}/{\tau}$ for a generic time-fractional single qubit open system versus the coupling strength $\lambda$ for four different driving times $\tau=0.1, 0.4, 0.7, 1$ and fractional orders $\beta=0.2, 0.5, 0.8, 1$ in four subfigures $(a)\sim(d)$. Parameter is set as $n=40$.}
\label{Fig3}
\end{figure}

We further study the influence of the non-Markovian memory effects of the environment on the QSL time of the time-fractional quantum dynamics. Fig. \ref{Fig3} shows a comparison of the QSL time ratio ${\tau _{QSL}}/{\tau}$ versus $\lambda$ in the cases of $\beta=0.2, 0.5, 0.8, 1$, $\tau=0.1, 0.4, 0.7, 1$ and $n=40$. From Fig. \ref{Fig3(a)} to Fig. \ref{Fig3(b)}, we observe for the same driving time, as small $\beta$ increases and exceeds a certain critical value, the non-Markovian oscillations emerge and enhance, which reveals the same result as the one from Fig. \ref{Fig2}: the smaller $\beta$ gets, the more the ability for the acceleration evolution in the time-fractional open quantum system will be. Besides, the short-driving-time dynamics shows no non-Markovian oscillations by viewing from Fig. \ref{Fig3(a)} to Fig. \ref{Fig3(d)}. For different driving times, in the non-Markovian region of the long-driving-time dynamics, its QSL time is smaller than that of the short-driving-time dynamics, as evidenced more strongly in Fig. \ref{Fig3(c)} and Fig. \ref{Fig3(d)}. This implies that the non-Markovian memory effects of the environment are able to speed up the time-fractional quantum evolution, resulting in a smaller QSL time. In particular, the Markovian to non-Markovian transition can be discovered at a long driving time when small $\beta$ exceeds a certain critical value, which presents a way to manipulate the non-Markovian dissipative dynamics of the time-fractional open quantum system.

\begin{figure}
\includegraphics[width=0.8\textwidth]{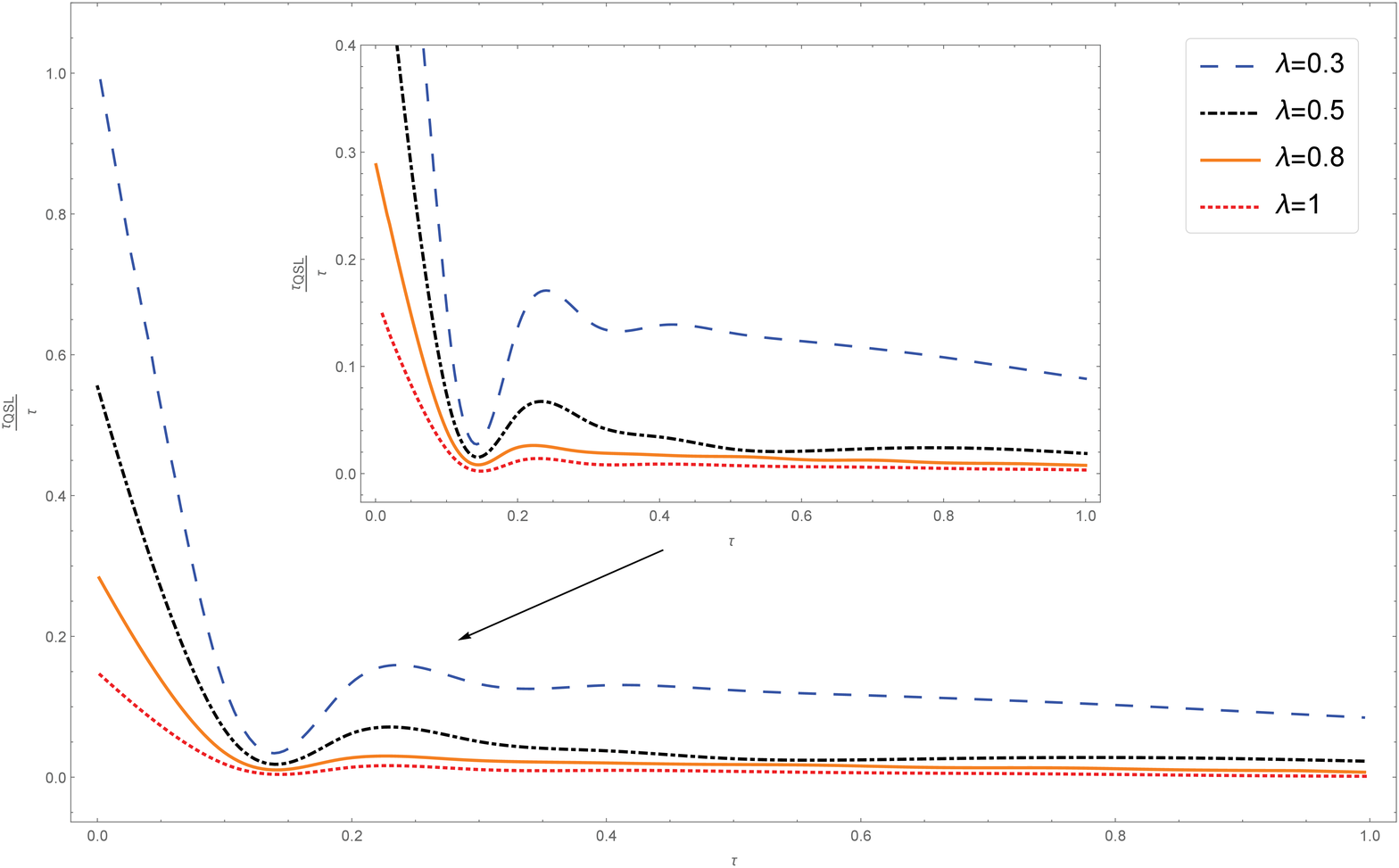}
\caption{The plot of ${\tau _{QSL}}/{\tau}$ for a generic time-fractional single qubit open system with different coupling strengths $\lambda=0.3, 0.5, 0.8, 1$ as a function of the driving time $\tau$. Settings of parameters are $\beta=0.5$, $n=20$. In the inset, we show how the small ${\tau _{QSL}}/{\tau}$ changes with different $\lambda$ for long driving times.}
\label{Fig4}
\end{figure}

In Fig. \ref{Fig4}, the parameters $\beta=0.5$ and $n=20$ are chosen for the plot of the QSL time ratio ${\tau _{QSL}}/{\tau}$ as a function of $\tau$ with $\lambda=0.3, 0.5, 0.8, 1$. We see that for short driving times, ${\tau _{QSL}}/{\tau}$ decays non-monotonically, while for long driving times, the non-Markovian oscillations disappear and no longer recur, which is the result of taking the small or intermediate $\beta$. Remarkably, for the long-driving-time behavior, all $\lambda$ are able to describe the non-Markovian features in the time-fractional open quantum system dynamics. In fact, for the smaller $\lambda$, the larger ${\tau _{QSL}}/{\tau}$ oscillates and decays more rapidly, pictured in the inset of Fig. \ref{Fig4}. Our analysis results demonstrate that the decrease of $\lambda$ inhibits the acceleration evolution of the time-fractional open quantum system, which is consistent with the result gained from the study on the traditional quantum system open dynamics in Ref. \cite{Teittinen2019}.

\begin{figure}[htbp]
\vspace{0.5cm}
\subfigtopskip=2pt
\subfigbottomskip=5pt
\subfigcapskip=-5pt
\centering
\subfigure[$\beta$=0.2]{
    \label{Fig5(a)}
    \includegraphics[width=0.46\linewidth]{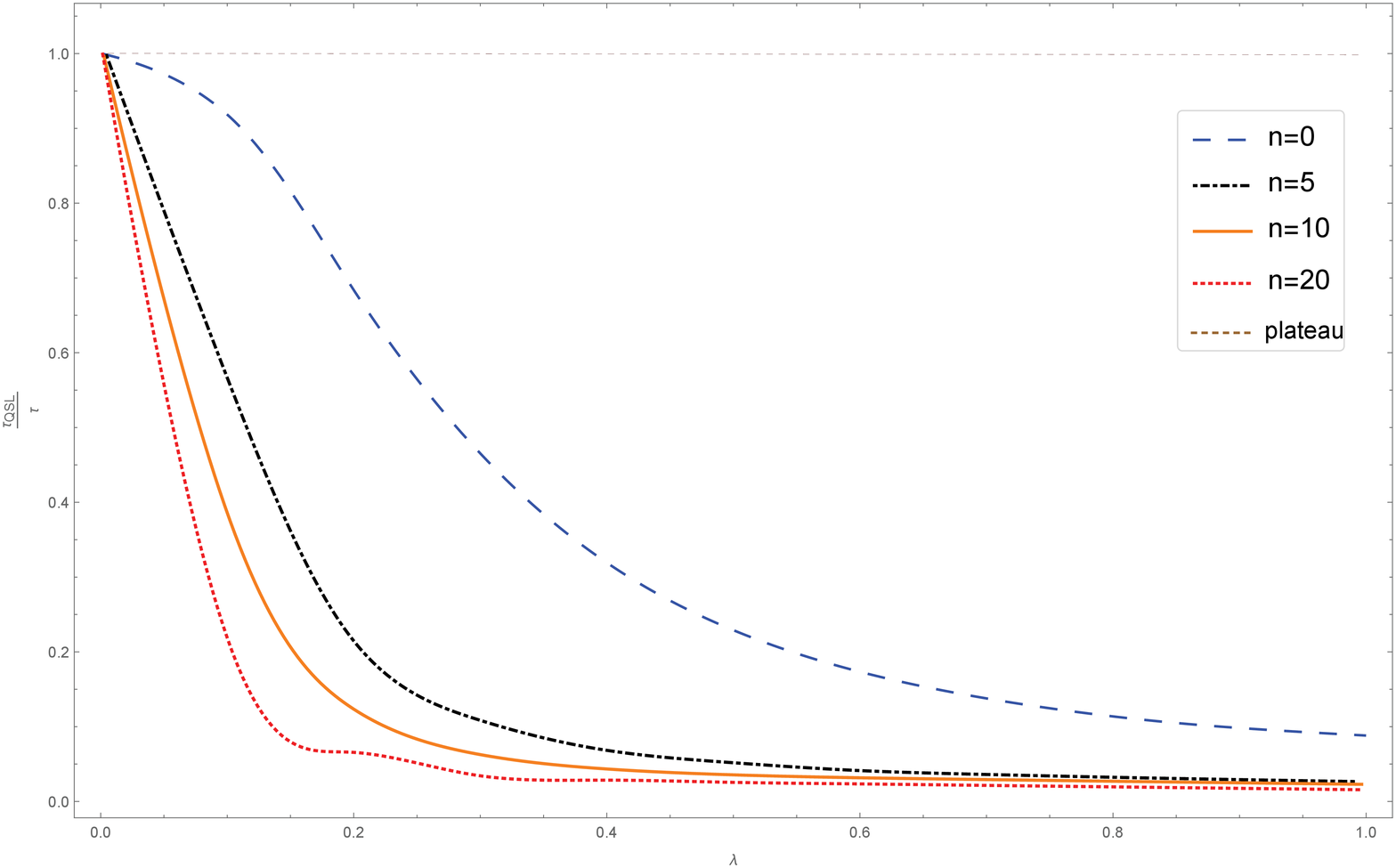}}
    \quad
\subfigure[$\beta$=0.5]{
    \label{Fig5(b)}
    \includegraphics[width=0.46\linewidth]{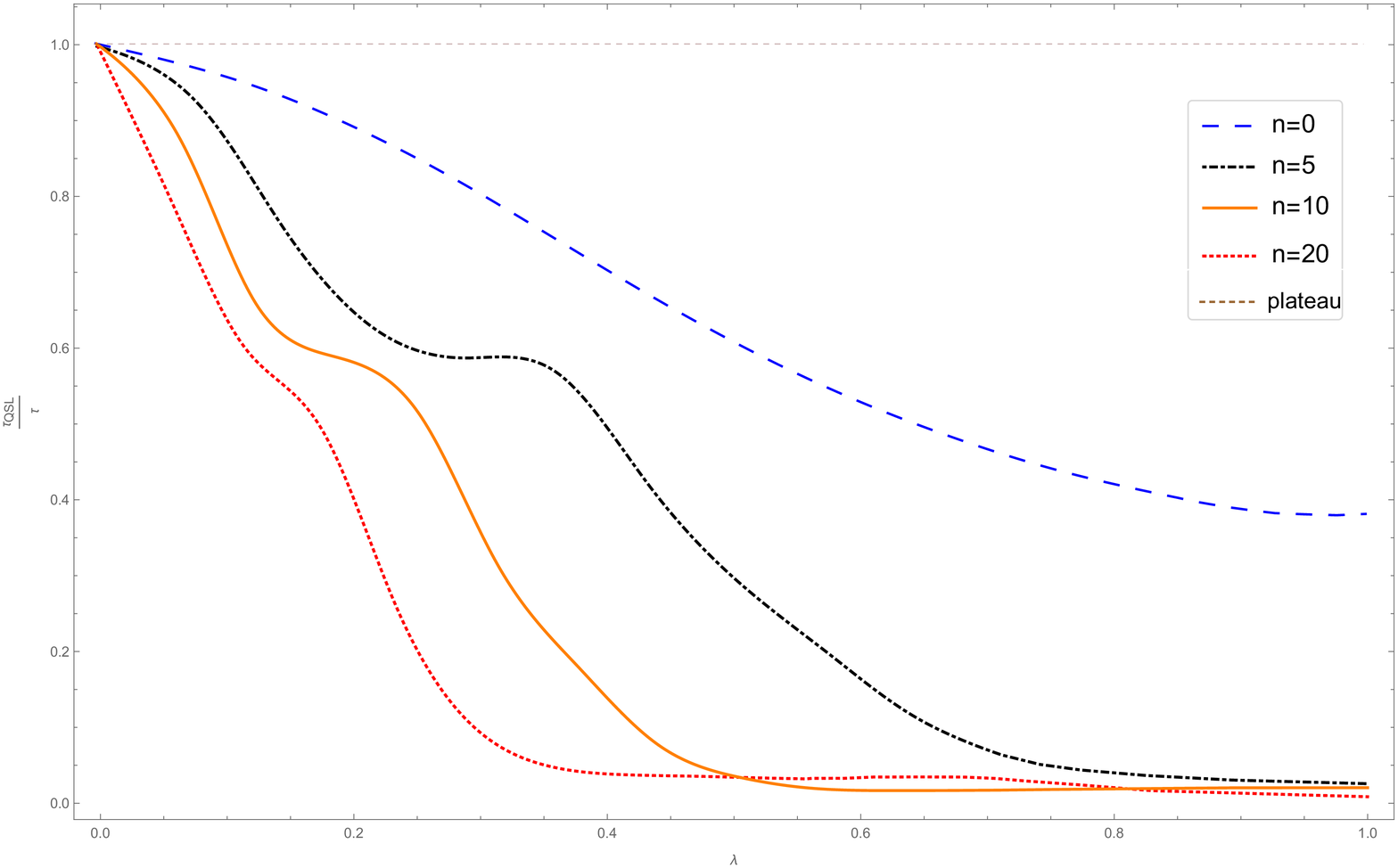}}\\
\subfigure[$\beta$=0.8]{
    \label{Fig5(c)}
    \includegraphics[width=0.46\linewidth]{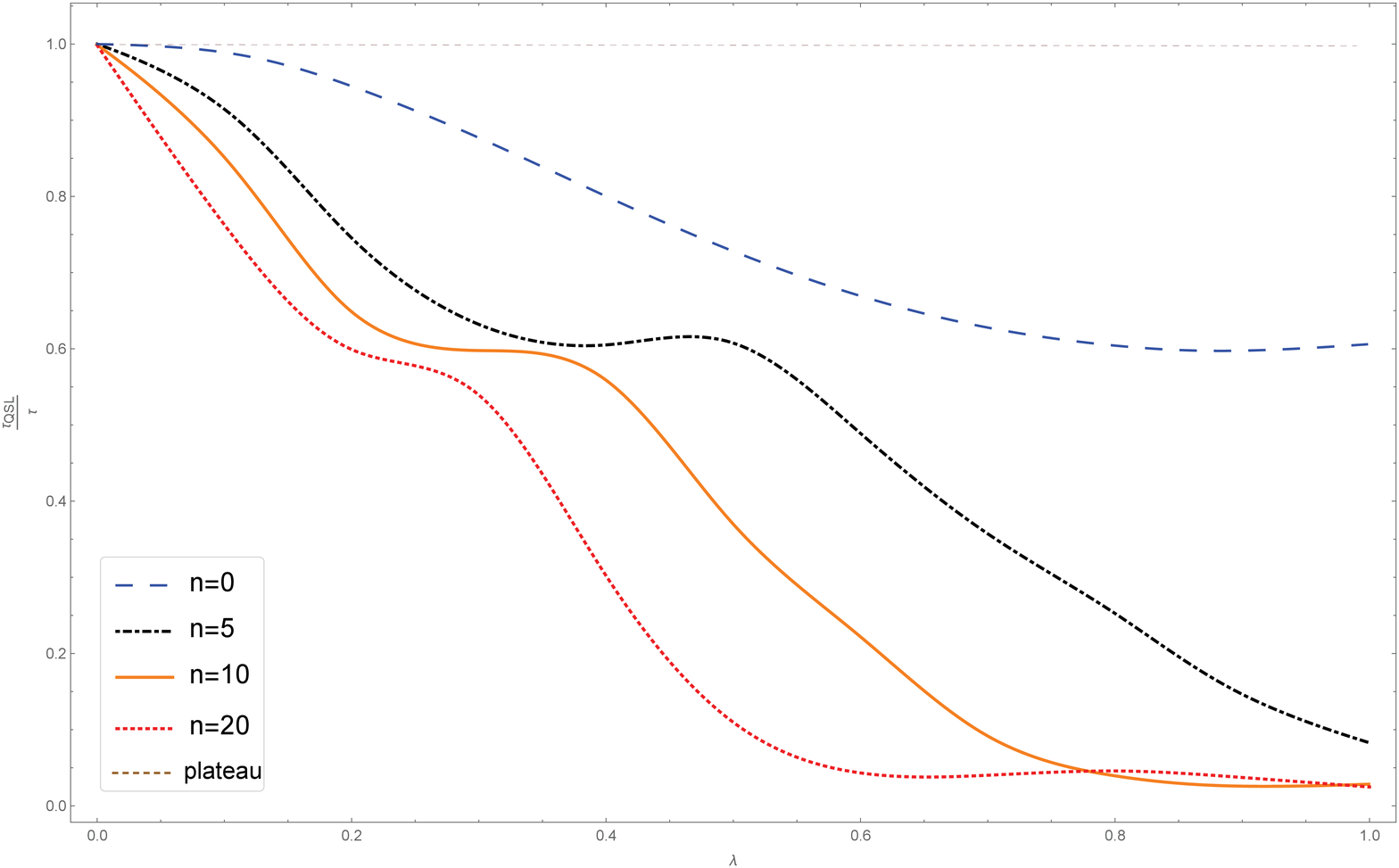}}
    \quad
\subfigure[$\beta$=1]{
    \label{Fig5(d)}
    \includegraphics[width=0.46\linewidth]{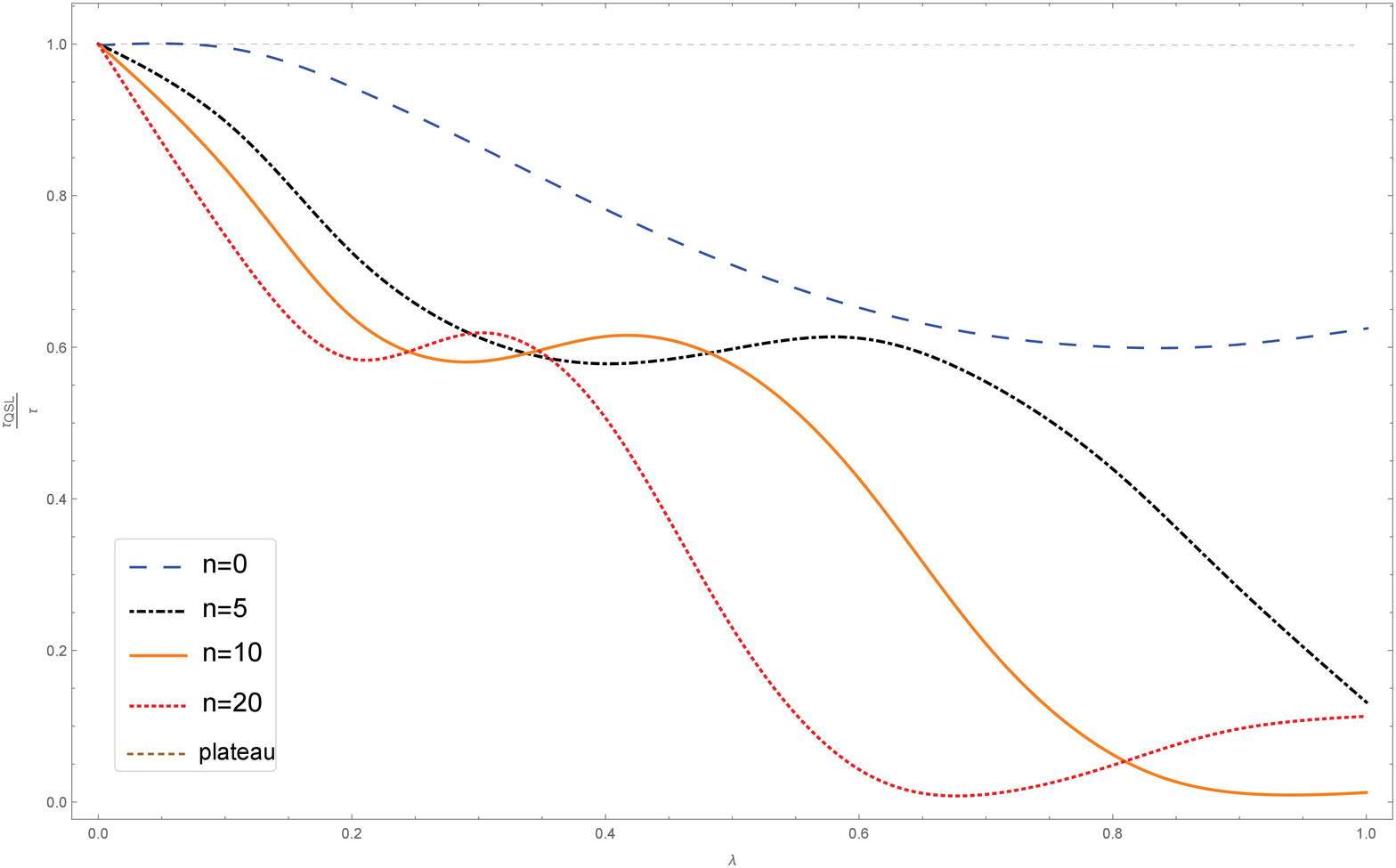}}\\
\caption{Comparison of ${\tau _{QSL}}/{\tau}$ for a generic time-fractional single qubit open system as a function of the coupling strength $\lambda$ between four different photon numbers $n=0, 5, 10, 20$ and fractional orders $\beta=0.2, 0.5, 0.8, 1$, as displayed by four subfigures $(a)\sim(d)$. Parameter is taken as $\tau=1$.}
\label{Fig5}
\end{figure}

To proceed with the generalization of our results, we now concentrate in the joint influence of $\lambda$, $n$ and $\beta$ on the QSL time for the time-fractional open quantum system under a given driving time, where the inquiry into the role of the non-Markovian memory effects is also indispensable. In Fig. \ref{Fig5} we present a comparison of the QSL time ratio ${\tau _{QSL}}/{\tau}$ as a function of $\lambda$, obtained from different $n=0, 5, 10, 20$ and $\beta=0.2, 0.5, 0.8, 1$ by fixing $\tau=1$. Clearly, when $n$ is very small, the non-Markovian oscillations become very inconspicuous. Moreover, the increase in $n$ may raise the oscillation frequency but reduce the amplitude, especially obvious at large enough $\beta$, like the ones shown in Fig. \ref{Fig5(c)} and Fig. \ref{Fig5(d)}. This indicates that the non-Markovian memory effects of the environment can result in a faster time-fractional quantum evolution, and thus to a smaller QSL time. On the whole, as $\beta$ decreases but both $\lambda$ and $n$ increase, the QSL time ratio ${\tau _{QSL}}/{\tau}$ decreases. That is to say, the smaller $\beta$ with larger $\lambda$ and $n$ can induce the more ability for further acceleration of the time-fractional quantum evolution in a memory environment.

There are the following three other meaningful results that can be drawn from the careful looks at the four subfigures in Fig. \ref{Fig5}:

(I) From Fig. \ref{Fig5(a)} to Fig. \ref{Fig5(b)}, for small $\beta$ above a certain critical value as well as small $\lambda$ and $n$, the transition point of ${\tau _{QSL}}/{\tau}$ from a monotonic function of $\lambda$ to a non-monotonic one is just the point at which the memoryless environment turns into the memory one. This coincides with the conditions found in Fig. \ref{Fig2} and Fig. \ref{Fig3} for the non-Markovian dissipative dynamics to take place.

(II) An interesting phenomenon about the plateau of ${\tau _{QSL}}/{\tau}=1$ arises in the plots indicated from Fig. \ref{Fig5(c)} to Fig. \ref{Fig5(d)}, it is explained by the direct dependence of the QSL time on the fractional order, coupling strength and photon number in Eq. (\ref{e22}): if the photon number and coupling strength are chosen very small but the fractional order very large for the dynamics to display no recoherence, in this case, the system can evolve along the fastest path, and hence ${\tau _{QSL}}/{\tau}=1$.

(III) A notable dynamical crossover from the no-acceleration to acceleration evolutions in the time-fractional open quantum system may appear at a certain critical value $\,{{\tilde \lambda}_{QSL}}$ of $\lambda$. When $\lambda\in\left[ {0,{{\tilde\lambda}_{QSL}}}\right]\,$, the evolution of the system is already the fastest, and has no ability for potential acceleration with increasing $\lambda$. And when $\lambda\in\left[{{{\tilde\lambda}_{QSL}},1}\right]\,$, the acceleration evolution of the system may occur and then the potential ability for further speedup increases as $\lambda$ increases.

In brief, we show that for a generic time-fractional single qubit open system, the non-Markovian memory effects of the environment play a crucial role in reducing the QSL time or improving the potential ability for further acceleration. Besides, the condition for the acceleration evolution in the time-fractional open system, that is, a tradeoff among the fractional order, coupling strength and photon number, is made clear. To be more specific, when the coupling strength and photon number are larger but the fractional order is smaller, the greater ability for potential acceleration of the time-fractional quantum evolution under noise will take place. Finally, it is valuable to state that the Markovian process can transit to the non-Markovian one as small fractional order is larger than a certain critical value at a long driving time, which offers a way to manipulate the non-Markovian dissipative dynamics of the time-fractional open quantum system by adjusting the fractional order.

\section{Conclusion}
\label{Sec:5}
In conclusion, we have explored the QSL time for a generic time-fractional single qubit open system by applying the TFSE to the resonant dissipative JC model. The results indicate that the non-Markovian memory effects of the environment are capable of speeding up the time-fractional quantum evolution, therefore resulting in a smaller QSL time. A linkage has been established between the QSL time and the fractional order, coupling strength along with photon number for a given driving time. With this, the condition for the speedup evolution in the time-fractional open quantum system, that is, a tradeoff among the three parameters, has been spelled out. Notably, the Markovian to non-Markovian transition can be controlled by adjusting the fractional order at a long driving time, which presents a method to manipulate the non-Markovian dissipative dynamics of a time-fractional open quantum system.

A procedure that can obtain the exact dynamics of a generic time-fractional single qubit interacting resonantly with its dissipative environment is shown. Our treatment can be directly extended to study the non-Markovian dissipative dynamics of the time-fractional multiqubit open systems, where each is locally coupled to its dissipative environment. This will take the idea forward for investigating the QSL time for the time-fractional multiqubit systems in uncorrelated and correlated composite environments \cite{Bellomo2007,Laine2012}. Our approach might be instructive for a deep understanding and description of the time-fractional dynamic evolution processes of quantum systems immersed in real environments.

On the other hand, the present work on the non-Markovian dissipative dynamics of open quantum systems is mainly focused on a basic two-level open system model, and its generalizations to the multi-level ones are not easy tasks. It is worth mentioning that tree-level systems (qutrits) have been put forward as a promising alternative to two-level systems (qubits) to be used in quantum processors \cite{Lanyon2008,Kumar2016}. Since quantum systems in reality always suffer from dissipation and decoherence, it will have important implications for further investigating the QSL time in the evolution of the time-fractional qutrit open system by considering a tree-level open system model, such as a V-type tree-level atom coupled to a dissipative reservoir \cite{Scully1997,Gu2012}.

\section*{Declaration of competing interest}
The authors declare that they have no known competing financial interests or personal relationships that could have appeared to influence the work reported in this paper.

\section*{CRediT authorship contribution statement}
\textbf{Dongmei Wei}: Conceptualization, Methodology, Formal analysis, Software, Data curation, Writing - Original draft, Writing - Review \& Editing, Visualization. \textbf{Hailing Liu}: Methodology, Formal analysis, Investigation, Writing - Review \& Editing. \textbf{Yongmei Li}: Methodology, Investigation, Data curation. \textbf{Fei Gao}: Supervision, Investigation, Project administration, Writing - Review \& Editing. \textbf{Sujuan Qin}: Supervision, Investigation, Project administration. \textbf{Qiaoyan Wen}: Project administration.

\section*{Acknowledgements}
This work is supported by National Natural Science Foundation of China (Grant Nos. 61976024, 61972048) and Beijing Natural Science Foundation (Grant No. 4222031).

\section*{Availability of data and materials }
The authors confirm that the data available for non-commercial using.

\end{document}